\documentclass[superscriptaddress,amsmath,twocolumn]{revtex4}
\usepackage{amsfonts}
\usepackage{graphicx}
\usepackage[latin1]{inputenc}
\usepackage{amsmath}
\usepackage{amssymb}

\begin{document}


\title{Mathisson-Papapetrou-Tulczyjew-Dixon equations in ultra-relativistic regime and gravimagnetic moment.}

\author{Alexei A. Deriglazov}
\email{alexei.deriglazov@ufjf.edu.br} \affiliation{Departamento de Matem\'atica, ICE, Universidade Federal de Juiz de
Fora, MG, Brazil} \affiliation{Laboratory of Mathematical Physics, Tomsk Polytechnic University, 634050 Tomsk, Lenin
Ave. 30, Russian Federation}
\author{Walberto Guzm\'an Ram\'irez }
\email{wguzman@cbpf.br} \affiliation{Departamento de Matem\'atica, ICE, Universidade Federal de Juiz de Fora, MG,
Brazil}

\date{\today}

\begin{abstract}
Mathisson-Papapetrou-Tulczyjew-Dixon (MPTD) equations in the Lagrangian formulation correspond to the minimal
interaction of spin with gravity. Due to the interaction, in the Lagrangian equations instead of the original metric
$g$ emerges spin-dependent effective metric $G=g+h(S)$. So we need to decide, which of them the MPTD particle sees as
the space-time metric. We show that MPTD equations, if considered with respect to original metric, have unsatisfactory
behavior: the acceleration in the direction of velocity grows up to infinity in the ultra-relativistic limit. If
considered with respect to $G$, the theory has no this problem. But the metric now depends on spin, so there is no
unique space-time manifold for the Universe of spinning particles: each particle probes his own three-dimensional
geometry. This can be improved by adding a non-minimal interaction of spin with gravity through gravimagnetic moment.
The modified MPTD equations with unit gravimagnetic moment have reasonable behavior within the original metric.
\end{abstract}

\maketitle


Equations of motion of a rotating body in a curved background are formulated usually in the multipole approach to
describe  the body \cite{Mathisson:1937zz, Papapetrou:1951pa, Tulc, Dixon1964, Dixon1965, Trautman2002, Costa2012,
Costa2014}. We consider the Mathisson-Papapetrou-Tulczyjew-Dixon (MPTD) equations, which describe the body in the
pole-dipole approximation, in the form studied by Dixon (for the relation of the Dixon equations with those of
Papapetrou and Tulczyjew see p. 335 in \cite{Dixon1964} as well as the recent works \cite{Costa2012, Costa2014}):

\begin{eqnarray}\label{gm1}
\nabla P^\mu=-\frac 14 R^\mu{}_{\nu\alpha\beta}S^{\alpha\beta}\dot x^\nu \equiv-\frac 14\theta^\mu{}_\nu\dot x^\nu, \cr
\nabla S^{\mu\nu}= 2P^{[\mu} \dot x^{\nu]}, \qquad S^{\mu\nu}P_\nu  =0,
\end{eqnarray}
(our $S^{\mu\nu}$ is twice that of Dixon). They are widely used now in computations of spin effects in compact binaries
and rotating black holes \cite{Thorne85, Kidder93, blanchet95, Khriplovich96, Pomeranskii1998, Khriplovich99,
Hinderer2013, Kunst14}, so our results may be relevant in this framework. In the multipole approach, $x^\mu(\tau)$ is
called the representative point (centroid) of the body, the antisymmetric spin-tensor $S^{\mu\nu}(\tau)$ is associated
with the inner angular momentum and the vector $P^\mu(\tau)$ is called momentum.

A few words about our notation. As we will discuss the ultra-relativistic limit, we do not assume the proper-time
parametrization, that means that we do not add the equation $g_{\mu\nu}\dot x^\mu\dot x^\nu=-c^2$ to the system (\ref{gm1}).
Our variables are taken in an arbitrary parametrization $\tau$, and $\dot x^\mu=\frac{dx^\mu}{d\tau}$. Then the
reparametrization-covariant system (\ref{gm1}) should be supplemented by a mass-shell equation, see \cite{DW2015}
for details. The covariant derivative is $\nabla P^\mu=\frac{dP^\mu}{d\tau} +\Gamma^\mu_{\alpha\beta}\dot x^\alpha P^\beta$
and the curvature is $R^\sigma{}_{\lambda\mu\nu}=\partial_\mu\Gamma^\sigma{}_{\lambda\nu} -\partial_\nu
\Gamma^\sigma{}_{\lambda\mu}+\Gamma^\sigma{}_{\beta\mu}\Gamma^{\beta}{}_{\lambda\nu}-
\Gamma^\sigma{}_{\beta\nu}\Gamma^{\beta}{}_{\lambda\mu}$. The square brackets mean antisymmetrization,
$\omega^{[\mu}\pi^{\nu]}=\omega^\mu\pi^\nu-\omega^\nu\pi^\mu$. We use the condensed notation $\dot x^\mu G_{\mu\nu}\dot
x^\nu=\dot xG\dot x$,
 $\omega^2=g_{\mu\nu}\omega^\mu\omega^\nu$, $\mu, \nu=0,
1, 2, 3$, $v^i\gamma_{ij}a^j={\bf v}\gamma{\bf a}$, $i, j=1, 2, 3$, and so on. The notation for the scalar functions
constructed from second-rank tensors are $\theta S= \theta^{\mu\nu}S_{\mu\nu}$, $S^2=S^{\mu\nu}S_{\mu\nu}$.

In the present work we continue the analysis of the MPTD equations on the base of Lagrangian formulation developed in the
recent work \cite{DW2015}, and discuss the necessity of generalizing  the formalism by  including an interaction of the spin with
gravity through non vanishing gravimagnetic moment. We discuss the behavior of the MPTD-particle in ultra-relativistic limit,
when the  speed of the particle approximates  the speed of light. Since we are interested in the influence of the spin on the particle's trajectory, we eliminate the momenta from the MPTD equations, thus obtaining a second-order equation for the
representative point $x^\mu(\tau)$. To achieve this, we compute the derivative of the spin supplementary condition (SSC),
$\nabla(S^{\mu\nu}P_\nu)=0$, and take into account that $P^2$ turns out to be constant of motion of the equations
(\ref{gm1}), say $\sqrt{-P^2}=k$, where $k/c$ can be called the mass of the particle \cite{Dixon1965}. Then the derivative implies \cite{DW2015}
\begin{eqnarray}\label{gm2}
P^\mu=\frac{k}{\sqrt{-\dot xG\dot x}}\tilde T^\mu_{\  \nu}\dot x^\nu, \qquad \tilde
T^\mu{}_\nu=\delta^\mu{}_\nu-\frac{1}{8k^2}S^{\mu\sigma}\theta_{\sigma\nu},
\end{eqnarray}
where  the matrix $G_{\mu\nu}$ is constructed from the "tetrad field" $\tilde T^\mu_{\  \nu}$ as follows:
\begin{equation}\label{gm3}
G_{\mu\nu}=g_{\mu\nu}+h_{\mu\nu}(S) \equiv g_{\alpha\beta} \tilde T^{\alpha}{}_{\mu} \tilde T^{\beta}{}_{\nu}.
\end{equation}
As this is composed from the original metric $g_{\mu\nu}$ plus the spin and field-dependent contribution $h_{\mu\nu}$,
we call $G_{\mu\nu}$ the effective metric produced {\it along the world-line} by the  interaction of spin with gravity.

Substituting  the expression (\ref{gm2}) into (\ref{gm1}) yields equations without $P^\mu$, modulo the constant of
motion $k$. Before we begin the analysis of the resulting equations, it is instructive to point out how they can be
obtained from a variational problem of the vector model of spin, see \cite{DW2015} for details.

Let us consider a relativistic spinning particle described by the  position $x^\mu(\tau)$ and by the vector $\omega^\mu(\tau)$
attached to the point $x^\mu$. The spin-tensor in our model is a composite quantity constructed from $\omega^\mu$ and
its conjugated momentum $\pi^\mu=\frac{\partial L}{\partial\dot\omega_\mu}$ as follows:
\begin{eqnarray}\label{gm4}
S^{\mu\nu}=2(\omega^\mu\pi^\nu-\omega^\nu\pi^\mu)=(S^{i0}=D^i, ~ S^{ij}=2\epsilon^{ijk}S^k).
\end{eqnarray}
Here $S^i$ is the three-dimensional spin-vector and $D^i$ is the dipole electric moment \cite{ba1}. The spinning particle in
the flat space is described by the Lagrangian action \cite{deriglazov2014Monster}
\begin{eqnarray}\label{gm5}
S=-\frac{1}{\sqrt{2}} \int d\tau \sqrt{m^2c^2 -\frac{\alpha}{\omega^2}} \qquad \qquad \quad \cr \times\sqrt{-\dot x N
\dot x - \dot\omega N \dot\omega +\sqrt{[\dot x N\dot x + \dot\omega N \dot\omega]^2- 4 (\dot x N \dot\omega )^2}}.
\end{eqnarray}
The matrix $N_{\mu\nu}=\eta_{\mu\nu}-\frac{\omega_\mu \omega_\nu}{\omega^2}$ is the projector on the  orthogonal plane
to $\omega^\nu$: $N_{\mu\nu} \omega^\nu=0$.  The double square-root structure in the expression (\ref{gm5}) seems to be
typical for the spin vector models \cite{hanson1974, mukunda1982, Hojman2013, Koch2016}. In the spinless limit, that is
$\alpha=0$ and $\omega^\mu=0$, the expression (\ref{gm5}) reduces to the standard Lagrangian of a relativistic
particle, $-mc\sqrt{-\dot x^\mu\dot x_\mu}$. Let us shortly enumerate some properties of the spinning particle.  The
Lagrangian depends on the free parameter $\alpha$ which determines the value of the  spin. The value
$\alpha=\frac{3\hbar^2}{4}$ corresponds to an elementary  one-half spin particle. The model has two local symmetries,
one is reparametrizations and the other is spin-plane symmetry  \cite{DPM1}. In the Hamiltonian formulation, the latter
symmetry acts on $\omega^\mu $ and $\pi^\nu$ but leaves $S^{\mu\nu}$ invariant, so only $S^{\mu\nu}$ is an observable
quantity. Canonical quantization of the model yields one-particle relativistic quantum mechanics with positive energy
states. For its relation with the Dirac equation, see \cite{DPM2}. At last, the model admits an interaction with
arbitrary electromagnetic \cite{DPM3} and gravitational \cite{DPW2} fields.

The minimal interaction with gravity is achieved by the covariantization of the formulation (\ref{gm5}), that is we
replace $\eta_{\mu\nu}\rightarrow g_{\mu\nu}$, and the usual derivative with the covariant one,
$\dot\omega^\mu\rightarrow \nabla\omega^\mu=\frac{d\omega^\mu}{d\tau} +\Gamma^\mu_{\alpha\beta}\dot
x^\alpha\omega^\beta$. For the further use, we also present the first-order (Hamiltonian) action of the theory
(\ref{gm4})
\begin{eqnarray}\label{gm7}
\int d\tau ~ p_\mu\dot x^\mu+\pi_\mu\dot \omega^\mu- \left[\frac{\lambda_1}{2}\left( P^2 + (mc)^2 + \pi^2 -
\frac{\alpha}{\omega^2}\right)\right. \cr\left. +\lambda_2 (\omega\pi) +\lambda_3 (P\omega)+\lambda_4 (P\pi)\right] .
\qquad \qquad \qquad
\end{eqnarray}
For the general-covariant phase-space quantities $\dot x^\mu$, $P_\mu=\frac{\partial L}{\partial\dot
x^\mu}-\Gamma^\beta_{\alpha\mu}\omega^\alpha\pi_\beta$ and $S^{\mu\nu}$, the variational problem yields the dynamical
equations (here $G_{\mu\nu}$ is given by (\ref{gm3}) with $\tilde
T^\mu{}_\nu=\delta^\mu{}_\nu-\frac{1}{8m^2c^2}S^{\mu\alpha}\theta_{\alpha\nu} $)
\begin{eqnarray}\label{gm6}
P^\mu =\frac{mc}{\sqrt{-\dot x G \dot x}}\tilde T^\mu_{\ \nu}\dot x^\nu, \qquad \nabla P^\mu =-\frac 14 \theta^\mu{}_\nu\dot
x^\nu \, , \cr \nabla S^{\mu\nu} = 2P^{[\mu} \dot x^{\nu]}\, ,  \qquad \qquad \qquad
\end{eqnarray}
along with the constraints
\begin{eqnarray}\label{gm6.1}
S^{\mu\nu}P_\nu  =0, \quad P^2+(mc)^2=0, \quad S^2=8\alpha \, ,
\end{eqnarray}
that is, the spin supplementary condition, the mass-shell equation and the value of the spin.

It is well known that $P^2=-k^2$ and $S^{\mu\nu}S_{\mu\nu}=\beta$ turn out to be constants of motion of the MPTD equations.
Comparing (\ref{gm6}) and (\ref{gm6.1}) with the MPTD equations (\ref{gm1}) and (\ref{gm2}), we conclude that all
trajectories of the MPTD equations with given integration constants $k$ and $\beta$ are described by our spinning particle
with mass $m=\frac{k}{c}$ and spin $\alpha=\frac{\beta}{8}$. Incidentally, we demonstrated that MPTD equations correspond
to the minimal interaction of the spinning particle with gravity.

Excluding the momenta $P^\mu$ from the system (\ref{gm6}) and (\ref{gm6.1}), we obtain the Lagrangian form of MPTD
equations

\begin{eqnarray}
\nabla\left[ \frac{ \tilde T^\mu{}_{\nu} \dot x^\nu}{\sqrt{-\dot xG\dot x}} \right] =-
\frac{1}{4mc}\theta^\mu{}_{\nu}\dot x^\nu \, , \label{gm8} \\
\nabla S^{\mu\nu}= \frac{1}{4mc\sqrt{-\dot xG\dot x}}\dot x^{[\mu} S^{\nu]\alpha}\theta_{\alpha\beta}\dot x^\beta \, , \label{gm9} \\
S^\mu{}_\nu\dot x^\nu-\frac{1}{8(mc)^2}S^{\mu}_{\  \alpha}S^{\alpha\beta}\theta_{\beta\nu}\dot x^\nu=0 \, . \label{gm10}
\end{eqnarray}
It is convenient to define the derivative $D$, $D\equiv\frac{1}{\sqrt{-\dot xG\dot x}}\frac{d}{d\tau}$. $D$ is the
reparametrization-invariant derivative: if $b(\tau)$ is a scalar under reparametrizations, $Db$ is a scalar as well.
Using $D$ instead of $\frac{d}{d\tau}$, Eq. (\ref{gm8}) can be presented in the form
\begin{eqnarray}\label{gm11}
DDx^\mu=f^\mu(Dx, S)\,.
\end{eqnarray}

All the subsequent discussion will be around the factor $\dot xG\dot x$, where the effective metric $G_{\mu\nu}$
appears. The equation for the trajectory (\ref{gm8}) becomes singular when the tangent vector $\dot x^\mu$ annihilates this
factor, $\dot xG\dot x=0$. The corresponding velocity  is called {\it critical velocity}. The observer independent
scale $c$ of special relativity is called, as usual, the speed of light. The singularity determines the behavior of the
particle in the ultra-relativistic limit. To clarify this point, two comments are in order: \par\noindent 1. Consider the
standard equations of a spinless particle in flat space interacting with an electromagnetic field in the physical-time
parametrization $x^\mu(t)=(ct, {\bf x}(t))$,
$\left(\frac{\dot x^\mu}{\sqrt{c^2-{\bf v}^2}}\right)^.=\frac{e}{mc^2}F^\mu{}_\nu \dot x^\nu$.
Then the factor is just $c^2-{\bf v}^2$, so the critical speed coincides with the speed of light. Rewriting the
equations in the form of Newton's Second Law, we find an acceleration. In this case, the longitudinal acceleration
reads $a_{||}={\bf v} {\bf a}=\frac{e(c^2-{\bf v}^2)^{\frac{3}{2}}}{mc^3}({\bf E}{\bf v})$. The factor, elevated in
some degree, appears on the right hand side of the equation and thus determines the value of the velocity at which the
longitudinal acceleration vanishes, $a_{||}\stackrel{v\rightarrow c}{\longrightarrow}0$. To resume, for the present
case the singularity implies \footnote{We point out that the factor can be hidden using the singular parametrization.
For instance, in the proper-time parametrization this would be encoded into the definition of $ds$, $ds=\sqrt{c^2-{\bf
v}^2}dt$.} that during its evolution in an electromagnetic field, the particle cannot exceed the speed of light $c$.
\par\noindent 2. In a curved space we have to be more careful since the three-dimensional geometry should respect the
coordinate independence of the speed of light. The notions for time interval, distance and velocity can be written
according to the following procedure \cite{Landau:2, deriglazovMPL2015}. For the events $x^\mu$ and $x^\mu+dx^\mu$ in a
curved space $g_{\mu\nu}$, the infinitesimal three-dimensional quantities are
\begin{eqnarray}\label{gm12}
dt=-\frac{g_{0\mu}dx^\mu}{c\sqrt{-g_{00}}}, \qquad \qquad \qquad \qquad \cr
dl^2=\left(g_{ij}-\frac{g_{0i}g_{0j}}{g_{00}}\right)dx^idx^j \equiv\gamma_{ij}(x^0, {\bf x})dx^idx^j.
\end{eqnarray}
Then the three-velocity vector ${\bf v}$ is
$v^i=\left(\frac{dt}{dx^0}\right)^{-1}\frac{dx^i}{dx^0}$ or, symbolically, $v^i=\frac{dx^i}{dt}$.
With these definitions, the four-interval acquires the form similar to special relativity: $g_{\mu\nu}dx^\mu
dx^\nu=-dt^2\left(c^2-{\bf v}\gamma{\bf v}\right)$, and a particle (photon) with propagation law $\dot xg\dot x=0$ has
the speed equal to $c$, $|{\bf v}\gamma{\bf v}|=c$.

To define an acceleration of a particle in the three-dimensional geometry, we need the notion of a constant vector
field (or, equivalently, the parallel-transport equation). The three-dimensional vector field $v^i$, given along a
curve $x^i(x^0)$, is called constant, if this obeys the following equation \cite{deriglazovMPL2015}: $\nabla_0
v^i+\frac12v^j(\partial_0\gamma_{jk})(\gamma^{-1})^{ki}=0$. Here the covariant derivative
$\nabla_0v^i=\frac{dv^i}{dx^0}+\tilde\Gamma^i{}_{jk}(\gamma)\frac{dx^j}{dx^0}v^k$ is defined with the help of the
three-dimensional metric $\gamma_{ij}(x^0,x^a)$, with $x^0$ considered as a parameter:
$\tilde\Gamma^i{}_{jk}(\gamma)=\frac12\gamma^{ia}(\partial_j\gamma_{ak}+\partial_k\gamma_{aj}-\partial_a\gamma_{jk})$.
This definition guarantees that  the scalar product of two constant fields does not depend on the point where it was
computed, $\frac{d}{dx^0}({\bf v}\gamma{\bf w})=0$. The deviation from a constant velocity is an acceleration
\begin{eqnarray}\label{gm13}
\mbox{} \quad a^i=\left(\frac{dt}{dx^0}\right)^{-1}\left[\nabla_0v^i+ \frac12
v^j(\partial_0\gamma_{jk})(\gamma^{-1})^{ki}\right].
\end{eqnarray}
The extra-term appearing  in this equation plays an essential role \cite{deriglazovMPL2015} in providing  that for the
geodesic motion we have: $a_{||}{\stackrel{v\rightarrow c}\longrightarrow} 0$. For the static metric,
$\partial_0\gamma_{ij}=0$, our definition reduces to that Landau-Lifshitz one, see page 251 in \cite{Landau:2}. Using the
expression (\ref{gm13}), we computed the longitudinal acceleration implied by the reparametrization-invariant equation
of a general form  (\ref{gm11})
\begin{eqnarray}\label{gm14}
{\bf v}\gamma{\bf a}=\frac{c^2-{\bf v}^2}{c^2} \left[(c^2-{\bf v}^2)({\bf v}\gamma{\bf f})+({\bf
v}\gamma)_j\tilde\Gamma^j{}_{ab}(\gamma)v^av^b \right. \cr \left. +\frac12(\frac{dt}{dx^0})^{-1} ({\bf
v}(\partial_0\gamma){\bf v})\right]. \qquad \qquad \qquad
\end{eqnarray}
This equation clearly shows that ultra-relativistic behavior is dictated by the number of relativistic factors
$(c^2-{\bf v}^2)^{-1}$ contained in the expression ${\bf v}\gamma{\bf f}$.

Let us return to the Lagrangian form (\ref{gm8})-(\ref{gm10}) of the MPTD equations. The singular factor contains the
effective metric $G_{\mu\nu}=g_{\mu\nu}+h_{\mu\nu}$ where $g_{\mu\nu}$ is the original metric. So we need to decide
which of them the particle sees as the space-time metric.

Let us use $g_{\mu\nu}$ to define the three-dimensional geometry (\ref{gm12}) and (\ref{gm13}). This leads us to two problems. The
first problem is that the critical speed turns out to be slightly more than the speed of light. To see this, we write
the equation which determines the critical speed
\begin{equation}\label{critical2}
\dot xG\dot x=0, ~  \mbox{or} ~   c^2-{\bf v}\gamma{\bf v}+ \frac{\left( \pi^2 (v\theta\omega)^2 +
\omega^2(v\theta\pi)^2 \right)}{(m^2c^2)^2}=0 \,.
\end{equation}
The surface is slightly different from the sphere $c^2-{\bf v}\gamma{\bf v}=0$. As $\pi_\mu$ and $\omega^\mu$ are space-like
vectors \cite{DPM3}, the last term is non-negative, this implies $|{\bf v}_{cr}|\ge c$. Let us confirm that generally
this term is nonvanishing function of velocity, then $|{\bf v}_{cr}|> c$ ( incidentally, this implies that $c$ does not
represent any special value for the equation (\ref{gm8})). Assuming  the contrary, that this term vanishes at some
velocity, then
\begin{eqnarray}\label{pe2}
v \theta \omega =\theta_{0i} \omega^i + \theta_{i0} v^i \omega^0 =0, \quad  v \theta \pi = \theta_{0i} \pi^i
+\theta_{i0} v^i \pi^0 =0 \, .
\end{eqnarray}
We analyze these equations in the following special case. We consider a space with covariantly-constant curvature
$\nabla_\sigma R_{\mu\nu\alpha\beta} =0$. Then $\frac{d}{d\tau}(\theta_{\mu\nu}S^{\mu\nu})=2\theta_{\mu\nu}\nabla
S^{\mu\nu}$, and using (\ref{gm9}) we conclude that $\theta_{\mu\nu}S^{\mu\nu}$ is an integral of motion. We further
assume that the only non vanishing is the electric \cite{Thorne85} part of the curvature, $R_{0i0j}=K_{ij}$. Then the
integral of motion acquires the form
%
$\theta_{\mu\nu} S^{\mu\nu} = 2 K_{ij}S^{0i}S^{0j}$.
%
Let us take the initial conditions for spin such that $K_{ij}S^{0i}S^{0j}\ne 0$, then this holds at any future instant.
Contrary to this,  the system (\ref{pe2}) implies $K_{ij}S^{0i}S^{0j}=0$. Thus, the critical speed does not always
coincide with the speed of light and, in general case, we expect that ${\bf v}_{cr}$ is both field and spin-dependent
quantity. The same conclusion follows from analytic and numeric computations on some specific backgrounds
\cite{Hojman2013, Koch2016}.

The second problem is that even in the case of a static field (and with $g_{0i}=0$) we obtained the rather surprising result that, due to
the factor $(-\dot x G\dot x)^{-\frac12}$, the longitudinal acceleration generally increases with velocity and becomes
infinite as the particle's speed approximates to the critical speed (the dots in Eq. (\ref{gm14.1}) state for
irrelevant to the present discussion non singular terms)
\begin{eqnarray}\label{gm14.1}
{\bf v}\gamma{\bf a}=\frac{1}{\sqrt{- vG v}} \frac{ v^i\gamma_{ij} T^j_{\ \alpha} S^{\alpha\sigma} R_{\sigma\nu\rho\lambda}v^\nu
v^\rho S^{\lambda\delta}\theta_{\delta \mu} v^\mu}{16 (mc)^3}+ \ldots \,,
\end{eqnarray}
where $T^\mu_{\ \nu}$ is the matrix inverse of $\tilde T^\mu_{\  \nu}$ in (\ref{gm2}). As $({\bf v}\gamma{\bf a})\sim
\frac{1}{m^3}$, this effect could be more appreciable for neutrinos. So MPTD equations, if considered with respect to
original metric, have unsatisfactory behavior in the ultra-relativistic limit.

Let us use $G_{\mu\nu}$ to define the three-dimensional geometry (\ref{gm12}) and (\ref{gm13}). In this case the expression for
longitudinal acceleration as a function of the force (\ref{gm11}) can be obtained in compact form for an arbitrary
original metric, and is written in Eq. (\ref{gm14}). From (\ref{gm8}) and (\ref{gm9}) we conclude that for the MPTD
particle ${\bf f}\sim(c^2-{\bf v}^2)^{-\frac{3}{2}}$, so the longitudinal acceleration (\ref{gm14}) vanishes as the
particle's speed approximates to the speed of light, $({\bf v}\gamma{\bf a})\stackrel{v\rightarrow
c}{\longrightarrow}0$. Since $G_{\mu\nu}$ is a  spin and field dependent quantity, we deduce that in this picture there is no
unique space-time manifold for the Universe of spinning particles, each particle will probe its own three-dimensional
geometry.

Can we modify the MPTD equations to obtain a theory with reasonable ultra-relativistic behavior  with respect to the
original metric $g_{\mu\nu}$? The effective metric $G_{\mu\nu}$ is spin and field dependent quantity, therefore, it
depends on the particular form of spin-field interaction. As we have seen above, the  MPTD equations result from a
minimal interaction of a spinning particle with a gravitational field. Adding a non minimal interaction, we have a
chance \footnote{Besides the SSC $S^{\mu\nu}P_\nu=0$, there are others SSC, see \cite{Costa2012, Costa2014} for a
review. By choosing another SSC, we modify the relation (\ref{gm2}) between $P$ and $\dot x$, and hence the equation
for the trajectory (\ref{gm8}). So changing the   SSC may be an alternative way to improve the ultra relativistic
behavior of the  MPTD particle.} to improve the ultra relativistic behavior of the MPTD particle. To achieve this, we
add the term $\frac{\lambda_1}{2}\frac{\kappa}{16} (\theta S) \equiv \frac{\lambda_1}{2}\kappa
R_{\alpha\beta\mu\nu}\omega^\alpha \pi^\beta \omega^\mu \pi^\nu$ into the Hamiltonian action (\ref{gm7}). This results
in the Hamiltonian variational problem (this can be compared with (\ref{gm7})
\begin{eqnarray}\label{gm14.2}
S_{\kappa}=\int d\tau ~ p_\mu\dot x^\mu+\pi_\mu\dot \omega^\mu-\qquad \qquad \cr [\frac{\lambda_1}{2}(
P^2+\frac{\kappa}{16}(\theta S) + (mc)^2 + \pi^2 - \frac{\alpha}{\omega^2})+\cr \lambda_2 (\omega\pi) +\lambda_3
(P\omega)+\lambda_4 (P\pi)]. \quad \qquad
\end{eqnarray}
In analogy with the magnetic moment, the interaction constant $\kappa$ is called gravimagnetic moment
\cite{Khriplovich1989, Pomeranskii}. For any value of $\kappa$, the new interaction turns out to be consistent with all
the constraints of the model. Equations of motion follow from variation of the action (\ref{gm14.2}), see
\cite{WGR2015} (approximate equations with nonvanishing gravimagnetic moment were discussed in \cite{Khriplovich1989,
Pomeranskii, Yee1992}). For the value $\kappa=0$,  the equations coincide with the initial MPTD equations. For a non
vanishing $\kappa$, the effective metric (\ref{gm3}) is constructed from the tetrad field which now depends on
$\kappa$, $\tilde T^{\alpha}{}_{\nu}= \delta^\alpha{}_\nu+(\kappa-1)bS^{\alpha \sigma}\theta_{\sigma\nu}$, where
$b=\frac{1}{8m^2c^2+\kappa(S\theta)}$. This can be compared with $\tilde T^\mu_{\ \nu}$ of the MPTD theory, see Eq.
(\ref{gm2}).

For the particular value $\kappa=1$, the effective metric $G_{\mu\nu}$ turns into the initial metric $g_{\mu\nu}$. If
we neglect unhomogeneous in curvature terms ($\nabla R=0$), the Hamiltonian equations read
\begin{eqnarray}
\nabla P_\mu =- \frac{1}{4}\theta_{\mu\nu}\dot x^\nu\, ,  \qquad \qquad \qquad \qquad \qquad \label{gmm191} \\
\nabla S^{\mu\nu} = -\frac{\sqrt{-\dot x g \dot x}}{4m_r c}(\theta S)^{[\mu\nu]}, \qquad S^{\mu\nu}P_\nu =0,
\label{gmm201}
\end{eqnarray}
where $P^\mu=\frac{m_r c}{\sqrt{-\dot x g \dot x}}\dot x^\mu$, and $m_r^2=m^2+\frac{(\theta S)}{16c^2}$. They can be
compared with MPTD equations (\ref{gm1}). We conclude that at small velocities our equation for precession of spin
differs from that of MPTD.

The Lagrangian equations read
\begin{eqnarray}
\nabla \left[ \frac{m_r\dot x^\mu}{\sqrt{-\dot xg\dot x}} \right] = -\frac{1}{4c} \theta ^\mu_{\ \nu} \dot x^\nu+\ldots \,, \label{im6} \\
\nabla S^{\mu\nu} =\frac{\sqrt{-\dot xg\dot x}}{4m_r c}\theta^{[\mu}_{ \  \alpha} S^{\nu]\alpha}+\ldots \,, \label{im7}
\end{eqnarray}
where the the dots signify the terms (which are non homogeneous in curvature, $O(\nabla R)$), that are not relevant to
the present discussion. These equations can be compared with (\ref{gm8}) and (\ref{gm9}). Even in a homogeneous field
we have modified dynamics for both $x$ and $S$. In the modified theory:
\par \noindent 1. The  time interval and the distance are unambiguously defined within the original space-time metric $g_{\mu\nu}$.
\par \noindent 2. The acceleration remains finite as $v\rightarrow c$, while the longitudinal acceleration vanishes as $v\rightarrow c$.
\par \noindent 3. The covariant precession of spin (\ref{im7}) has a smooth behavior, in
particular, for homogeneous field, $\nabla R=0$, we have $\nabla S\sim \sqrt{-\dot x g\dot x}$ contrary to $\nabla
S\sim\frac{1}{\sqrt{-\dot x g\dot x}}$ in the equation (\ref{gm9}).

That means, in contrast with MPTD equations, the modified theory is consistent with respect to the original metric
$g_{\mu\nu}$. Hence the modified equations could be more promising for description of the rotating objects in
astrophysics.

In conclusion, we note that the  MPTD equations follow from a particular form assumed for the multipole representation
of a rotating body \cite{Trautman2002}. It would be interesting to find a set of multipoles which yields the modified
equations (\ref{im6}) and (\ref{im7}).

\section*{Acknowledgments}
The work of AAD has been supported by the Brazilian foundations CNPq (Conselho Nacional de
Desenvolvimento Cient\'ifico e Tecnol\'ogico - Brasil) and FAPEMIG (Fundac\~ao de Amparo \'a Pesquisa do Estado de Minas Gerais
- Brasil). WGR thanks CAPES for the financial support (Program PNPD/2011).


\end{document}